\documentclass[floatfix,showpacs,showkeys,aps,prl,twocolumn]{revtex4-2}
\usepackage{lineno}
\usepackage{amsmath}
\usepackage{amssymb}

\usepackage{tikz}
\usepackage{lipsum,adjustbox}
\usepackage{graphicx}
\usepackage[caption = false]{subfig}
\usepackage{dcolumn}
\usepackage{bm}
\usepackage{gensymb}
\usepackage{multirow}
\usepackage{tabstackengine}
\usepackage[inline]{enumitem}

\setstackEOL{\cr}
\setstackgap{L}{\normalbaselineskip}

\let\oldequation\equation
\let\oldendequation\endequation

\renewenvironment{equation}
{\linenomathNonumbers\oldequation}
{\oldendequation\endlinenomath}

\usepackage{hyperref}
\hypersetup{
	pdftitle={},
	pdfauthor={},
	pdfsubject={},
	pdfkeywords={},
	pdfstartview={},
	bookmarksopen=true, breaklinks=true, debug=true,
	colorlinks=true, linkcolor=blue, citecolor=blue, urlcolor=blue,
	hyperfigures=true
}

\usepackage{todonotes}

\newcommand{\pilnu}{D \to \pi\ell^+\nu_{\ell}}
\newcommand{\pienu}{D^0\to \pi^-e^+\nu_e}
\newcommand{\pimunu}{D^0\to \pi^-\mu^+\nu_\mu}
\newcommand{\pizenu}{D^+\to \pi^0e^+\nu_e}
\newcommand{\pizmunu}{D^+\to \pi^0\mu^+\nu_\mu}

\newcommand{\ffpi}{f^{D\to\pi}_+(0)}

\newcommand{\lfudz}{0.947 \pm 0.014_{\rm stat.} \pm 0.005_{\rm syst.}}
\newcommand{\lfudp}{0.973 \pm 0.015_{\rm stat.} \pm 0.006_{\rm syst.}}

\newcommand{\pwre}{1.025 \pm 0.013_{\rm stat.} \pm 0.007_{\rm syst.}}
\newcommand{\pwrmu}{1.011 \pm 0.019_{\rm stat.} \pm 0.010_{\rm syst.}}

\newcommand{\fpi}{0.6339\pm0.0024_{\rm stat.}\pm0.0014_{\rm syst.}}
\newcommand{\vcdpi}{0.2262\pm0.0008_{\rm stat.}\pm0.0005_{\rm syst.}\pm0.0018_{\rm LQCD.}}

\newcommand{\BESIIIorcid}[1]{\href{https://orcid.org/#1}{\hspace*{0.1em}\raisebox{-0.45ex}{\includegraphics[width=1em]{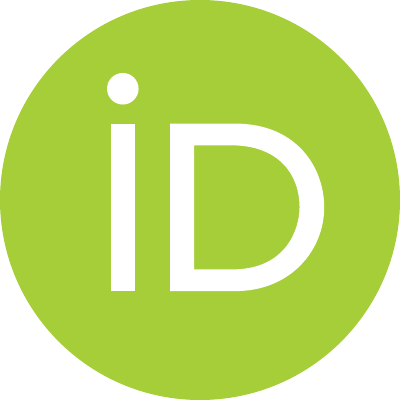}}}}

\begin{document}

\title{\bf \boldmath Precision Measurement of Decay Dynamics in $D^{0(+)}\to \pi^{-(0)}\ell^+\nu_\ell$}

\author{M.~Ablikim$^{1}$\BESIIIorcid{0000-0002-3935-619X},
M.~N.~Achasov$^{4,c}$\BESIIIorcid{0000-0002-9400-8622},
P.~Adlarson$^{83}$\BESIIIorcid{0000-0001-6280-3851},
X.~C.~Ai$^{89}$\BESIIIorcid{0000-0003-3856-2415},
C.~S.~Akondi$^{31A,31B}$\BESIIIorcid{0000-0001-6303-5217},
R.~Aliberti$^{39}$\BESIIIorcid{0000-0003-3500-4012},
A.~Amoroso$^{82A,82C}$\BESIIIorcid{0000-0002-3095-8610},
Q.~An$^{78,65,\dagger}$,
Y.~H.~An$^{89}$\BESIIIorcid{0009-0008-3419-0849},
M.~S.~Anderson$^{39}$\BESIIIorcid{0009-0008-1550-2632}
% ...
\\
\vspace{0.2cm}
(BESIII Collaboration)\\
\vspace{0.2cm} {\it
$^{1}$ Institute of High Energy Physics, Beijing 100049, People's Republic of China\\
% ...
$^{4}$ Budker Institute of Nuclear Physics SB RAS (BINP), Novosibirsk 630090, Russia\\
% ...
$^{31}$ INFN Sezione di Ferrara, (A)INFN Sezione di Ferrara, I-44122, Ferrara, Italy; (B)University of Ferrara, I-44122, Ferrara, Italy\\
% ...
$^{39}$ Johannes Gutenberg University of Mainz, Johann-Joachim-Becher-Weg 45, D-55099 Mainz, Germany\\
% ...
$^{65}$ State Key Laboratory of Particle Detection and Electronics, Beijing 100049, Hefei 230026, People's Republic of China\\
% ...
$^{78}$ University of Science and Technology of China, Hefei 230026, People's Republic of China\\
% ...
$^{82}$ University of Turin and INFN, (A)University of Turin, I-10125, Turin, Italy; (B)University of Eastern Piedmont, I-15121, Alessandria, Italy; (C)INFN, I-10125, Turin, Italy\\
$^{83}$ Uppsala University, Box 516, SE-75120 Uppsala, Sweden\\
% ...
$^{89}$ Zhengzhou University, Zhengzhou 450001, People's Republic of China\\
\vspace{0.2cm}
$^{\dagger}$ Deceased\\
% ...
$^{c}$ Also at the Novosibirsk State University, Novosibirsk, 630090, Russia\\
% ...
}
}

\begin{abstract}
	The branching fractions of $D^0\to \pi^-e^+\nu_e$, $D^0\to \pi^-\mu^+\nu_\mu$, $D^+\to \pi^0e^+\nu_e$, and $D^+\to \pi^0\mu^+\nu_\mu$ are precisely measured, using 20.3 fb$^{-1}$ of $e^+e^-$ collision data collected at the center-of-mass energy of 3.773 GeV with the BESIII detector. The ratios of the decay widths between muon and positron channels are examined in full, across several four-momentum transfer ranges of $\ell^+\nu_{\ell}$. No lepton flavor universality violation is found in the current data.
	From a simultaneous fit to the precisely measured partial decay rates and the first measured forward-backward asymmetries of these four decays, the product of the hadronic transition form factor, $f^{D\to\pi}_+(0)$, and the modulus of the $c\to d$ quark mixing element, $|V_{cd}|$, is measured with unprecedented precision to be $f^{D\to\pi}_+(0)|V_{cd}|=0.1425\pm0.0005_{\rm stat.}\pm0.0003_{\rm syst.}$. Taking the value of $|V_{cd}|$ from the standard model global fit and $f^{D\to\pi}_+(0)$ derived by the lattice quantum chromodynamics calculation as input, we obtain $f^{D\to\pi}_+(0)=0.1425\pm0.0005_{\rm stat.}\pm0.0003_{\rm syst.}$ and $|V_{cd}|=0.2262\pm0.0008_{\rm stat.}\pm0.0005_{\rm syst.}\pm0.0018_{\rm LQCD.}$, respectively. The precision of each result is a factor of 2-3 better than the previous best measurements. Additionally, the real and imaginary parts of the scalar current contribution in the $c\to d \ell^+\nu_{\ell}$ transition are measured for the first time to be Re $(C_S^\mu)=$ $0.022 \pm 0.023_{\rm stat.}\pm 0.003_{\rm syst.}$ and $|\mathrm{Im} (C_S^\mu)|=0.000 \pm 0.038_{\rm stat.}\pm 0.012_{\rm syst.}$.
\end{abstract}

\maketitle

Experimental studies of semileptonic (SL) decays of heavy mesons are critical for testing the Standard Model (SM), and probing potential new physics (NP) effects beyond the SM. In the SM, the couplings of different families of leptons to gauge bosons are expected to be equal. Recent experiments from BaBar, Belle, and LHCb~\cite{BaBar:2012obs,BaBar:2013mob,Belle:2015qfa,Belle:2016ure,LHCb:2015gmp} reported anomalies in SL $B$ decays mediated by $b \to c \ell^+\bar{\nu}_{\ell}$ that deviate from the SM predictions up to 3 $\sigma$~\cite{Li:2018lxi,Bifani:2018zmi}, indicating a hint of the violation of lepton flavor universality (LFU). 
Precision measurements of the branching fraction (BF) ratios and decay dynamics of SL $D$ decays would offer important complementary information to understand these tensions. 

Various NP models, including two-Higgs-doublet models~\cite{Iguro:2022uzz,Blanke:2022pjy}, leptoquark scenarios~\cite{Sakaki:2013bfa,Becirevic:2018afm}, and model-independent investigations~\cite{Iguro:2024hyk}, suggest that the violation of LFU may arise from a scalar current  (SC) contribution in weak interaction.
Investigating the $c\to s(d)\ell^+\nu_\ell$ transitions in SL $D$ decays is crucial for understanding these discrepancies~\cite{Fajfer:2015ixa,Leng:2020fei,Jain:2025kqe}.
In the $W_{\ell^+\nu_{\ell}}$ rest frame, $\theta_W$ denotes the angle between the lepton momentum and the direction opposite to the $D$ meson. The partial decay rates (PDRs) $\frac{d\Gamma}{dq^2}$, the forward $(\cos\theta_{W}>0)$ and backward $(\cos\theta_{W}<0)$ asymmetries $\mathcal{A}_{FB}(q^2)$ of $D\to \pi \ell^+\nu_\ell$ are dependent on a SC contribution in the $c\to d\ell^+\nu_\ell$ transition, making these studies essential for constraining the SC parameter space and potentially revealing unknown scalar bosons beyond the SM.

Accurate measurements of the decay dynamics of $\pilnu$ also allow for  precise extraction of the product $|V_{cd}|f^{D\to\pi}_+(0)$, where $|V_{cd}|$  is the Cabibbo-Kobayashi-Maskawa (CKM) matrix
element and $f^{D\to\pi}_+(0)$ is the hadronic transition form factor (FF) at zero-momentum transfer.
The hadronic FF can be calculated using various theoretical approaches, such as the quark model, QCD sum rules, and lattice quantum chromodynamics (LQCD)~\cite{Melikhov:2000yu,Verma:2011yw,Soni:2018adu,Ivanov:2019nqd,Faustov:2019mqr,Khodjamirian:2000ds,Wang:2002zba,Wu:2006rd,FermilabLattice:2004ncd,Na:2011mc,Lubicz:2017syv,FermilabLattice:2022gku}.
The precisely measured values of $f^{D\to\pi}_+(0)$ and $|V_{cd}|$ are important for testing different theoretical predictions and verifying the unitarity  of the CKM matrix with higher precision.

Previously, several BF measurements and decay dynamics analyses of $D^0 \to \pi^-\ell^+\nu_\ell$ and $D^+ \to \pi^0\ell^+\nu_\ell$ were reported by BaBar~\cite{BaBar:2014xzf}, CLEO-c ~\cite{CLEO:2005cuk,CLEO:2005rxg,CLEO:2007ntr,CLEO:2009svp}, Belle~\cite{Belle:2006idb} and BESIII~\cite{BESIII:2015tql,BESIII:2017ylw,BESIII:2018nzb}.
However, no experimental studies of the SC and the forward-backward asymmetry were reported. 
This Letter reports the first experimental constraints on the SC in the $D\to \pi\ell^+\nu_\ell$ decays and the precisely determined PDRs.
The BFs, the hadronic FF $f^{D\to \pi}_+(0)$, and the modulus of the CKM  matrix element $|V_{cd}|$, as well as 
tests of LFU in full and several $q^2$ ($=(E_{\ell^{+}}+E_{\nu_{\ell}})^2-(\vec{p}_{\ell^{+}}+\vec{p}_{\nu_{\ell}})^2$) intervals,  are reported based on analysis of 20.3~fb$^{-1}$~\cite{Ablikim:2013ntc} of $e^+e^-$ collision data collected by the BESIII detector at the center-of-mass energy $\sqrt s=3.773$ GeV.
These results supersede the previous BESIII measurements~\cite{BESIII:2015tql,BESIII:2017ylw,BESIII:2018nzb} that used a portion of the dataset presented here.
Throughout this Letter, charge conjugated modes are implied unless stated otherwise.

The BESIII detector~\cite{BESIII:2009fln,BESIII:2020nme} consists of a helium-based multilayer drift chamber (MDC), a plastic scintillator time-of-flight system (TOF), and a  CsI(Tl)  electromagnetic calorimeter (EMC), all enclosed in a superconducting solenoidal magnet providing a 1.0 T magnetic field. The inclusive Monte Carlo (MC) sample described in Refs.~\cite{geant4,kkmc,conexc,evtgen,lundcharm,photos,Tchikilev:1999vpw} has been validated for background simulation.
The signal MC events of $\pilnu$ are generated using the two-parameter series expansion~\cite{Tchikilev:1999vpw}, with the parameters obtained in this Letter.

The $\psi(3770)$ resonance decays mainly into a $D\bar{D}$ pair. Throughout this Letter, $D$ refers to $D^0(D^+)$ and $\bar{D}$ refers to $\bar{D}^0(D^-)$ unless stated explicitly. 
If a single-tag (ST) $\bar{D}$ meson is correctly reconstructed, the presence of another $D$ meson is guaranteed.
Thus, in the system recoiling against an ST $\bar{D}$ meson, the double-tag (DT) candidates for $D^{0(+)}\to \pi^{-(0)}\ell^+\nu_\ell$ are selected.
The ST $\bar{D}$ mesons are reconstructed using decay modes:
$\bar{D}^0\to K^+\pi^-$, $K^+\pi^-\pi^0$, $K^+\pi^-\pi^-\pi^+$, $K^+\pi^-\pi^0\pi^0$ and $K^+\pi^-\pi^-\pi^+\pi^0$; $D^-\to K^{+}\pi^{-}\pi^{-}$,
$K^0_{S}\pi^{-}$, $K^{+}\pi^{-}\pi^{-}\pi^{0}$, $K^0_{S}\pi^{-}\pi^{0}$, $K^0_{S}\pi^{+}\pi^{-}\pi^{-}$
and $K^{+}K^{-}\pi^{-}$. 
The BF of the $\pilnu$ decay is determined by
\begin{equation}
	\label{eq:bf}
	{\mathcal B} = \frac{N_{\rm DT}}{N^{\rm tot}_{\rm ST}\bar{\varepsilon}_{\rm sig}},
\end{equation}
where $N^{\rm tot}_{\rm ST} = \sum_{i}N^{i}_{\rm ST}$ and $N_{\rm DT}$ are the ST and DT yields in data,
$\bar{\varepsilon}_{\rm sig}=\sum_{i}[(\varepsilon^{i}_{\rm DT}N^{i}_{\rm ST})/(\varepsilon^{i}_{\rm ST}N^{\rm tot}_{\rm ST})]$ represents the efficiency of finding the $D^{0(+)}\to \pi^{-(0)}\ell^+\nu_\ell$ decay in the presence of an ST $\bar D$ meson.

The selection criteria of $K^\pm$, $\pi^\pm$, $K^0_S$, $\pi^0$ meson candidates and photon candidates on both tag and signal sides are the same as those in Ref.~\cite{BESIII:2024slx}.
The ST $\bar D$ mesons are selected based on two kinematic variables defined in the $e^+e^-$ center-of-mass frame: the energy difference $\Delta E\equiv E_{\bar D}-E_{\mathrm{beam}}$ and the beam-constrained mass $M_{\rm BC}\equiv\sqrt{E_{\mathrm{beam}}^{2}/c^{4}-|\vec{p}_{\bar D}|^{2}/c^{2}}$. Here, $E_{\mathrm{beam}}$ is the beam energy, and $E_{\bar D}$ and $\vec{p}_{\bar D}$ are the total energy and momentum of the $\bar D$ candidate, respectively. 
If there are multiple combinations for a given ST mode, the one giving the minimum $|\Delta E|$ is retained.

For each tag mode, the ST yield and efficiency are extracted by an unbinned maximum likelihood fit to the $M_{\rm BC}$ distribution of the reconstructed $\bar D$ candidates from data and  inclusive MC sample, respectively. 
The fit method is the same as those in Ref.~\cite{BESIII:2024slx}. The $\Delta E$ requirements, the ST yields in data, and the ST efficiencies are provided in Ref.~\cite{supplement}.
Summing over all tag modes, the total yields of ST $D^-$ and $\bar D^0$  mesons are $(19,057.6 \pm 5.4_{\rm stat.})\times 10^3$ and $(10,646.9\pm3.8_{\rm stat.})\times 10^3$, respectively.

The $\pilnu$ candidates are reconstructed from the remaining tracks and showers that are not used in the tag side.
Candidates for $\pi^{-}$ and $\pi^0$ mesons are selected using the same criteria as those employed in the ST selection.
To ensure data-MC consistency and minimize systematic uncertainties, lepton candidates are required to satisfy $|\cos\theta|<0.80$. 
They are identified by using the ionization energy loss measured by the MDC, along with information from the TOF and EMC, based on which the likelihoods for positron, muon, kaon and pion hypotheses $\mathcal{L}_{e,\mu,\pi,K}$ are calculated.
Positron candidates must satisfy $\mathcal{L}_e /(\mathcal{L}_e+\mathcal{L}_\pi+\mathcal{L}_K) > 0.8$, and $\mathcal{L}_e >0.001$. 
In addition,
their deposited energies in the EMC are required to be greater than 0.8 times their momenta reconstructed by the MDC.
The four-momenta of nearby photons are aligned with the positron momentum to partially account for energy loss due to final-state radiation (FSR).
Muon candidates must satisfy $\mathcal{L}_\mu > \mathcal{L}_e$, and $\mathcal{L}_\mu >0.001$, with energies deposited in the EMC being less than 0.3 GeV.
The muons with momentum greater than 0.5 GeV/$c$ are required to satisfy the
criterion of 
momentum-dependent hit depth in the muon counter as used in Ref.~\cite{BESIII:2018nzb}.
To suppress the backgrounds from the hadronic $D$ decays involving $\pi^0$ meson in final states, there must be no additional good charged tracks in the signal side ($N_{\rm extra}^{\rm char}$) and the maximum energies of extra photons ($E_{\text{extra~}\gamma}^{\rm max}$) not used in the event selection have to be less than 0.25 GeV.

To suppress the backgrounds associated with the misidentification between $\pi^{+}$ and $\ell^{+}$ particles, the invariant masses of the $\pi^{-(0)}\ell^+$ combinations are required to be less than 1.80 (1.70) GeV/$c^2$ for the positron (muon) channels.
To further reject the peaking backgrounds from $D^0\to K_{S}^0(\pi^{+}\pi^{-})\pi^0$ and $D^{+}\to\bar{K}^{0}\pi^+$ decays for the signal modes $\pimunu$ and $\pizmunu$, we require $M_{\pi^{-}\mu^{+}}$ and $M_{D^{-}\mu{+}}^{\rm rec}$ ($D^{-}\mu^{+}$ recoil mass) to be outside the ranges (0.455, 0.497) GeV/$c^2$ and (0.437, 0.580) GeV/$c^2$, respectively. These ranges correspond to 
the regions within 
$\pm 3\sigma$ around each fitted peak.

The undetected neutrino is inferred from the kinematic variable $M_{\rm miss}^{2} \equiv E_{\rm miss}^{2}/c^{2} - |\vec{p}_{\rm miss}|^{2}/c^{2}$, which peaks around zero for signal events. 
The $E_{\rm miss}$ and $\vec{p}_{\rm miss}$, given by $E_{\rm miss}\equiv E_{\rm beam}-E_{\pi}-E_{\ell^+}$ and $\vec{p}_{\rm miss}\equiv\vec{p}_{D}-\vec{p}_{\pi}-\vec{p}_{\ell^+}$, are the missing energy and momentum of the DT event in the $e^+e^-$ center-of-mass frame, respectively. Here, $E_{\pi(\ell^+)}$ and $\vec{p}_{\pi(\ell^+)}$ are the measured energy and momentum of the $\pi(\ell^+)$ candidates, respectively, and $\vec{p}_{D}\equiv-\hat{p}_{\bar D} \sqrt{E_{\rm beam}^2/c^2-m_{\bar D}^2 c^2 }$, where $\hat{p}_{\bar D}$ is the unit vector in the momentum direction of the ST $\bar D$ meson and $m_{\bar D}$ is the nominal $\bar D$ mass~\cite{pdg2024}.

\begin{figure}[htbp]
	\centering
	\includegraphics[width=0.98\columnwidth]{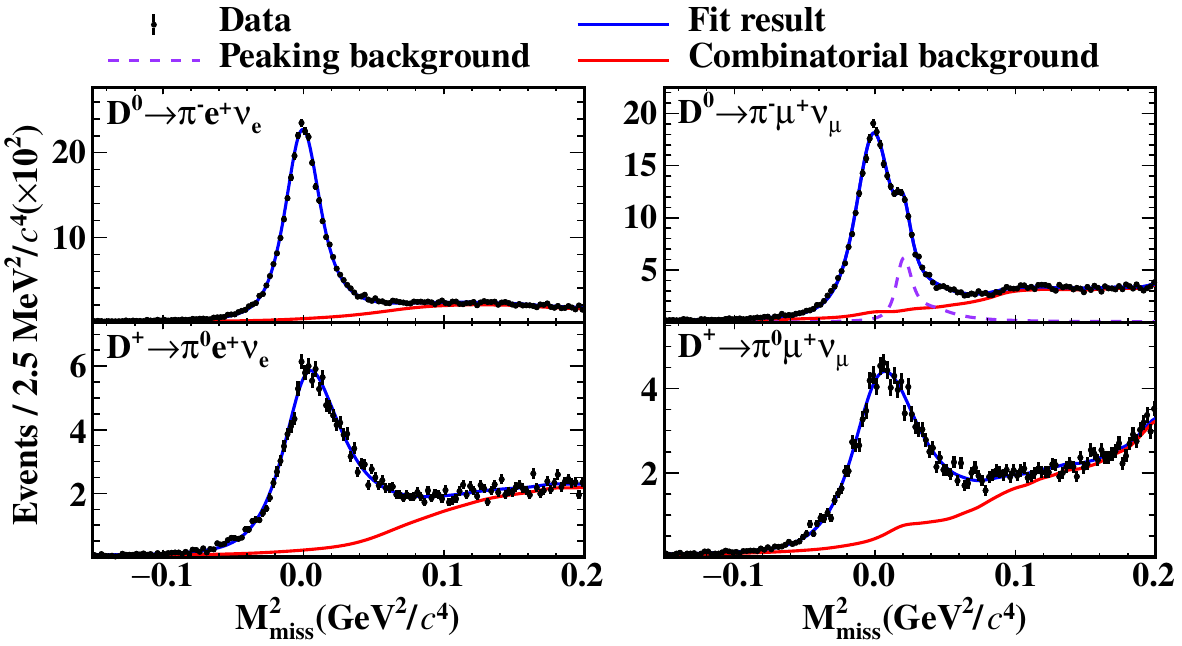}
	\caption{Fits to the $M_{\rm miss}^2$ distributions for different signal decays.}
	\label{fig:fitmm2m}
\end{figure}

After imposing all selection criteria, the $M_{\rm miss}^{2}$ distributions are shown in Fig.~\ref{fig:fitmm2m}. 
Fits to these distributions are performed  to extract the signal yields.
The signal and background shapes are derived from MC simulation, with their yields floated. The signal shape is convolved with a Gaussian function with free parameters to account for resolution difference between data and MC simulation.
The peaking background $D^{0}\to\pi^{-}\pi^{+}\pi^{0}$ for the channel $\pimunu$ is described by its simulated shape convolved with a different Gaussian function.

The BF of each signal decay is determined using the signal yields in data~($N_{\rm DT}$), the weighted signal efficiencies~($\bar \varepsilon_{\rm sig}$) corrected by the data-MC discrepancies, and ST $\bar D$ yields in data ($N^{\rm tot}_{\rm ST}$). The results are summarized in Table~\ref{table:bfsum}.

\begin{table}[htbp]
	\caption{Signal yields in data ($N_{\rm DT}$), averaged signal efficiencies ($\bar{\varepsilon}_{\rm sig}$), and obtained BF for each signal decay ($\mathcal B_{\rm sig}$), where the first uncertainties are statistical and the second systematic.}
	\label{table:bfsum}
	\centering
		\begin{tabular}{lccc}
			\hline\hline
			Decay                             & $\bar{\varepsilon}_{\rm sig}(\%)$  & $N_{\rm DT}$      & $\mathcal{B}_{\rm sig}(\times10^{-3})$ \\ \hline
			$D^{0}\to\pi^{-}e^{+}\nu_{e}$     & $61.97$                     & $34834\pm202$ & $2.950\pm0.017\pm0.017$              \\
			$D^{0}\to\pi^{-}\mu^{+}\nu_{\mu}$ & $46.84$                     & $25140\pm330$ & $2.817\pm0.037\pm0.019$              \\
			$D^{+}\to\pi^{0}e^{+}\nu_{e}$     & $39.76$                     & $15330\pm145$ & $3.622\pm0.034\pm0.018$              \\
			$D^{+}\to\pi^{0}\mu^{+}\nu_{\mu}$ & $29.52$                     & $10890\pm135$ & $3.507\pm0.043\pm0.026$              \\
			\hline\hline
		\end{tabular}
\end{table}

The systematic uncertainties in BF measurements are discussed below.
The uncertainties of ST $\bar D^0$ and $D^-$ yields in data, $N_{\rm ST}^{\rm tot}$, are both assigned as 0.30\% by varying the signal and background shapes in the $M_{\rm BC}$ fits.
The uncertainties for tracking and particle identification (PID) of $e^+$ (0.20\%), $\mu^+$(0.20\%), $\pi^-$(0.20\%), and $\pi^0$ reconstruction (0.17\% for $e^+$ channels, 0.19\% for $\mu^+$ channels) are investigated using control samples of $e^+e^-\to\gamma e^+ e^-$, $e^+e^-\to\gamma \mu^+ \mu^-$, and hadronic DT $D\bar D$ events, respectively. 
The systematic uncertainties from $E_{\rm extra\gamma}^{\rm max}$ and $N^{\rm extra}_{\rm char}$ requirements (0.10\%) are investigated using the control sample of hadronic DT $D\bar D$ events.
The uncertainties of signal MC model (0.03\%$\sim$0.13\%) are estimated by comparing DT efficiencies obtained by varying the input parameters by $\pm1\sigma$.
The uncertainty of the $K_{S}^{0}$ veto (0.06\% and 0.02\%) is estimated by changing the fitted resolution based on MC simulation. 
The systematic uncertainty for the $M_{\rm miss}^{2}$ fit is estimated to be (0.06\%$\sim$0.54\%) by varying the relative fractions of the backgrounds from $e^+e^- \to q\bar{q}$ and the dominant $D$ background channels.
The uncertainties of limited MC statistics (0.02\%$\sim$0.03\%), FSR effect of the positron channels (0.26\% and 0.10\%), and the BF of $\pi^0\to\gamma\gamma$ (0.03\%) are also included.
The difference in measured BFs between the simultaneous and nominal fit methods (using single yield-weighted efficiencies) is taken as a systematic uncertainty (0.10\% and 0.10\%).
Adding all systematic uncertainties in quadrature gives the total systematic uncertainty for each signal decay. They are 0.58\%, 0.66\%, 0.50\%, and 0.73\% for the $\pienu$, $\pimunu$, $\pizenu$, and $\pizmunu$ channels, respectively.

With the measured BFs, we determine the BF ratios between muon and positron channels to be
$\mathcal{R}_{\rm LFU}^{0} =  \lfudz$ and
$\mathcal{R}_{\rm LFU}^{+} = \lfudp$ for the $D^0$ and $D^+$ meson decay separately.
Here, the systematic uncertainties in ST yields, $\pi^-/\ell^+$ tracking and PID, and $\pi^0$ reconstruction cancel. The results are consistent with the SM prediction of $0.985 \pm 0.002$~\cite{Riggio:2017zwh} within 2.5$\sigma$ and 0.8$\sigma$, respectively.
When combining the $D$ meson lifetimes $\tau_{D^0} = (410.3~\pm~1.0)$ fs and $\tau_{D^+} = (1033~\pm~5)$ fs~\cite{pdg2024},
we determine the BF ratios between $D^0$ and $D^+$ channels, ($\mathcal{R}_{\rm IS}^{\ell} \equiv \frac{\Gamma(D^{0}\to\pi\ell^+\nu_{\ell})}{2\Gamma(D^{+}\to\pi\ell^+\nu_{\ell})}$), to be
$\mathcal{R}_{\rm IS}^{e} = \pwre$ and $\mathcal{R}_{\rm IS}^{\mu} = \pwrmu$. 
The systematic uncertainties cancel similarly to the $\mathcal{R}_{\rm LFU}$ calculation, 
while those from the lifetimes are added. These ratios support isospin conservation in these decays.

To study the decay dynamics of $\pilnu$, candidate events are divided into different $q^2$ intervals with a step size of 0.3 GeV$^2/c^4$. The $q^2$-binned PDRs $\Delta\Gamma_i^{\rm msr}$ are determined as $\Delta\Gamma^{\rm msr}_{i}=\frac{N_{\rm pro}^{i}}{\tau_{D}\cdot N_{\rm ST}^{\rm tot}}$, where $\tau_{D}$ is the $D$ meson lifetime~\cite{pdg2024} and $N_{\rm pro}^{i}=\sum_j^{N_{\rm bins}}\left(\varepsilon^{-1}\right)_{ij}N_{\rm DT}^{j}$ is the corresponding produced signal SL yield. Here, the observed SL yield ($N_{\rm DT}^{j}$) is obtained by fitting the corresponding $M_{\rm miss}^{2}$ distribution in the $j$-th reconstructed $q^{2}$ interval. 
The $\varepsilon^{-1}$ is the inverse matrix of the signal efficiency matrix $\varepsilon$, which includes the migrations between the generated and reconstructed $q^{2}$ intervals choice.
In addition to the systematic uncertainties in the BF measurement, two additional uncertainties arise from $\tau_{D}$~\cite{pdg2024} and the different $q^{2}$ interval choice. Detailed information of correlated and uncorrelated systematic uncertainties are given in Ref.~\cite{supplement}.

The $q^{2}$-binned forward-backward asymmetry in the $i$th $q^{2}$ interval is defined as
\begin{equation}
	\small
	\mathcal{A}_{FB,i}=\int_{q_{\text{min}(i)}^{2}}^{q_{\text{max}(i)}^{2}}\mathcal{A}_{FB}(q^{2})\frac{d\Gamma}{dq^{2}}dq^{2}\left/\int_{q_{\text{min}(i)}^{2}}^{q_{\text{max}(i)}^{2}}\frac{d\Gamma}{dq^{2}}dq^{2}.\right.
\end{equation}
They are measured with 
\begin{equation}
	\small
	\mathcal{A}_{FB,i}^{\rm msr} = \frac{N_{\rm pro}(\cos\theta_{W}>0,q^2_i)-N_{\rm pro}(\cos\theta_{W}<0,q^2_i)}{N_{\rm pro}(\cos\theta_{W}>0,q^2_i)+N_{\rm pro}(\cos\theta_{W}<0,q^2_i)},
\end{equation}
similarly to the PDRs. Except those from $\pi^-$ tracking and PID ($\pi^0$ reconstruction), $\ell^+$ tracking and PID, $M_{\rm miss}^2$ fit, and MC statistics, other systematic uncertainties of $N_{\rm pro}$ cancel in the calculation of $\mathcal{A}_{FB,i}$. They are estimated similarly to the PDRs.

\begin{figure*}[htbp]
	\centering
		\includegraphics[width=\linewidth]{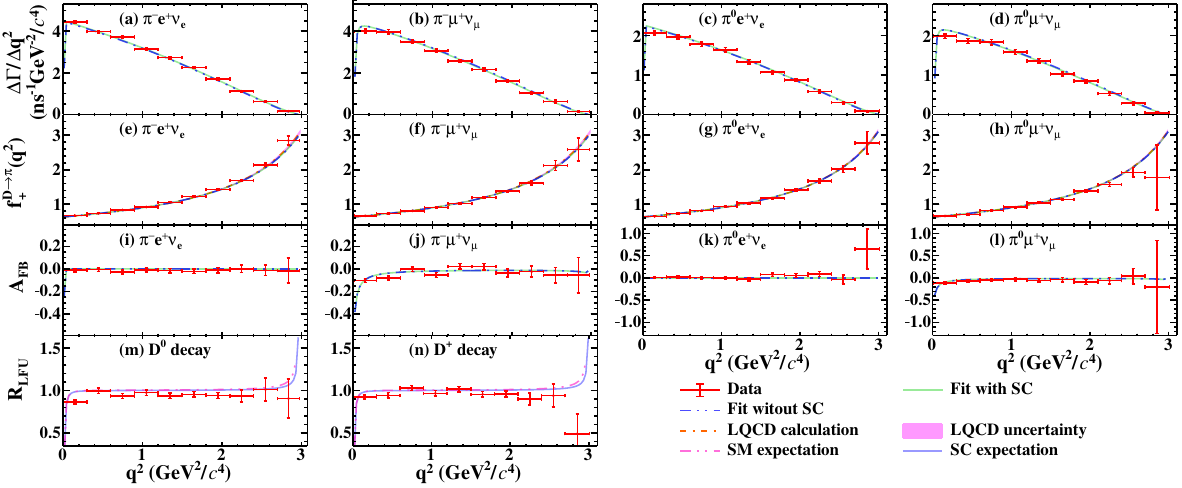}
		\caption{Simultaneous fit to the measured $\frac{d\Gamma}{dq^2}$ (a,b,c,d) and $\mathcal{A}_{FB}(q^2)$ (i,j,k,l), the projections on $f^{D\to \pi}_+(q^2)$ (e,f,g,h), the measured and theoretical expectation $\mathcal{R}_{\rm LFU}$ (m,n).}
		\label{ff_pilnu}
\end{figure*}

The measured PDRs of $\pilnu$ and the calculated $\mathcal R_{\rm LFU}$ in each $q^2$ interval are shown in Figs.~\ref{ff_pilnu}(a-h, m, n). The determined $\mathcal{A}_{FB}(q^2)$ for $\pilnu$ are shown in Figs.~\ref{ff_pilnu} (i-l).

The theoretical decay rates $\Delta\Gamma_i^{\rm th}=\int_{q^{2}_{\rm min, (i)}}^{q^{2}_{\rm max, (i)}}\frac{d\Gamma}{dq^2}dq^2$ and forward-backward asymmetries $\mathcal{A}_{FB}^{\rm th}(q^2)$ are expressed as~\cite{Faustov:2019mqr}
\begin{equation}
	\scriptsize
	\frac{d \Gamma}{d q^{2}} = X \mathcal{N}(q^{2})\left(1-\frac{m_{\ell}^{2}}{q^{2}}\right)^{2}\left[(1+\frac{m_{\ell}^{2}}{2q^{2}})\left|\mathcal{H}_{0}(q^{2})\right|^{2}+\frac{3 m_{\ell}^{2}}{2 q^{2}}\left|\mathcal{H}_{t}(q^{2})\right|^{2}\right],
	\label{eq:pwr}
\end{equation}
and 
\begin{equation}
	\scriptsize
	\mathcal{A}_{FB}^{\rm th}(q^2) = \frac{3 \mathcal{N}(q^{2})}{2} \frac{1}{d \Gamma / d q^{2}}\left(1-\frac{m_{\ell}^{2}}{q^{2}}\right)^{2} \frac{m_{\ell}^{2}}{q^{2}}{\rm Re}\left(\mathcal{H}_{0}(q^{2}) \mathcal{H}_{t}(q^{2})\right).
\end{equation}
The term $\mathcal{N}(q^2) = \frac{G_{F}^{2}|V_{cd}|^{2}|\mathbf{q}|q^2}{96\pi^{3}m_{D}^{2}}$ is the overall normalization factor, incorporating the Fermi coupling constant $G_F$, the CKM matrix element $|V_{cd}|$, and the $D$ meson mass $m_D$~\cite{pdg2024}. 
Here, $\mathbf{q}$ denotes the four-momentum of the $\ell^+\nu_{\ell}$ system in the $D$ meson rest frame, and $m_\ell^+$ is the lepton mass.
In Eq.~\ref{eq:pwr}, $X$ is a multiplicative isospin factor, which is 1 for $D^{0}\to\pi^{-}\ell^{+}\nu_{\ell}$ and $1/2$ for $D^{+}\to\pi^{0}\ell^{+}\nu_{\ell}$.

The hadronic helicity amplitudes $\mathcal{H}_{0(t)}(q^2)$ are expressed as
\begin{equation}
	\footnotesize
	\mathcal{H}_{0}(q^2)=\frac{2m_D|\mathbf{q}|}{\sqrt{q^2}}f^{D\to\pi}_+(q^2),
	\mathcal{H}_{t}(q^2)=\frac{m_D^2-m_\pi^2}{\sqrt{q^2}}f^\pi_0(q^2),
	\label{equation:heliamp}
\end{equation}
and with SC contributions, $\mathcal{H}_{t}(q^2)$ is modified as
\begin{equation}
	\small
	\mathcal{H}_{t}(q^2)= \left(1+C_S^\ell\frac{q^2}{m_\ell(m_d-m_c)}\right)\frac{m_D^2-m_\pi^2}{\sqrt{q^2}}f^{D\to\pi}_0(q^2).
	\label{equation:heliampm}
\end{equation}
Here, $f^{D\to\pi}_{+}(q^{2})$ and $f^{D\to\pi}_{0}(q^{2})$ are the vector and scalar FFs, respectively. The $C_S^{\ell}=C_R^{\ell}+C_L^{\ell}$ represents the scalar combination of Wilson coefficients for the right and left-handed SCs~\cite{Fajfer:2015ixa,Leng:2020fei,Jain:2025kqe}. The values $m_c = 1.273$ GeV$/c^2$ and $m_d = 4.67$ MeV$/c^2$~\cite{pdg2024} are the $\overline{\rm MS}$ masses at the energy scale $\mu=m_s$ and 2 GeV, respectively; $m_\pi$ is the pion mass. 
The Wilson coefficient $C_{S}^{\ell}$ can be complex in both positron and muon channels. For the muon channels, angular observables allow $\operatorname{Re}(C_S^\mu)$ and $\operatorname{Im}(C_S^\mu)$ to be treated separately. For the positron channels, the PDRs are only sensitive to $|C_S^e|$, and forward-backward asymmetry is not affected by the scalar contribution; thus only $|C_S^e|$ is considered.

The hadronic FFs $f^{D\to\pi}_{+}(q^{2})$ are parameterized using a two-parameter series expansion~\cite{Becher:2005bg}, which is a function of the product $f^{D\to\pi}_{+}(0)|V_{cd}|$ and the shape parameter $r_{1}$.
For $f^{D\to\pi}_{0}(0)$, a one-parameter expansion is used due to its much smaller contribution.

We perform a simultaneous fit to the differential decay rates and the forward-backward asymmetries of $\pilnu$ by minimizing 
\begin{equation}
	\chi^2=\chi_{\Delta\Gamma}^{2}+\sum\chi_{\mathcal{A}_{\rm FB}}^{2}.
\end{equation}
Here, the $\chi^2_{\Delta\Gamma}$ sums over all four signal decays while accounting for correlations between different $q^2$ intervals and signal decays using the total covariance matrix. For the $\sum\chi^{2}_{\mathcal{A}_{\rm FB}}$, correlations between different signal decays are ignored  since the only correlated uncertainty arises from correction factors of $\ell^{+}$, which is limited. The $\chi^2_{\Omega}$ is expressed as 
\begin{equation}
	\small
	\chi^2_{\Omega} = \sum_{i,j}\left(\Omega_{i}^{\rm msr}-\Omega_{i}^{\rm th}\right)\times C_{\Omega,ij}^{-1}\left(\Omega_{j}^{\rm msr}-\Omega_{j}^{\rm th}\right),
\end{equation}
where $\Omega$ denotes $\Delta\Gamma$ or $\mathcal{A}_{\rm FB}$ and $C_{ij}$ is the sum of the statistical and systematic covariance matrices.
Two separate fits are conducted under the hypotheses with and without a SC contribution. The results of the fit to the PDRs and the projections on $f^{D\to \pi}_+(q^2)$ are shown in the Figs.~\ref{ff_pilnu}(a-h); and the obtained fit results are summarized in Table~\ref{table:afbpar}. 
Since LFU may not apply to the SC, depending on lepton mass, its muon coupling could exceed its positron coupling.
Through a fit where $C_S^\mu$ is allowed to vary and 
$|C_S^e|$ is fixed to zero, the significance of non-zero $C_S^\mu$ is evaluated to be 1.6 $\sigma$ relative to the null hypothesis (without SC).

\begin{table}[htbp]
	\centering
	\caption{Parameters obtained from the fit with and without the SC, where the first uncertainties are statistical and the second systematic. The ndf denotes the number of degrees of freedom. }
	\label{table:afbpar}
		\begin{tabular}{lll}
			\hline\hline
			Variable                     & With SC             & Without SC     \\
			\hline
			$f^{D\to\pi}_{+}(0)|V_{cd}|$ & 0.1430(06)(04)      & 0.1425(05)(03) \\
			$r_{1}$                      & -1.97(03)(01)       & -1.99(03)(01)  \\
			$|C_S^e|$                    & 0.059(13)(09)       & ...            \\
			${\rm Re}(C_S^\mu)$          & 0.022(23)(03)       & ...            \\
			${\rm Im}(C_S^\mu)$          & 0.00(38)(12)        & ...            \\
			$\chi^2/\rm ndf$             & $75.5/75$           & $81.0/78$      \\
			\hline
			\hline
		\end{tabular}
\end{table}

\begin{figure}[htbp]
	\centering
	\includegraphics[width=\columnwidth]{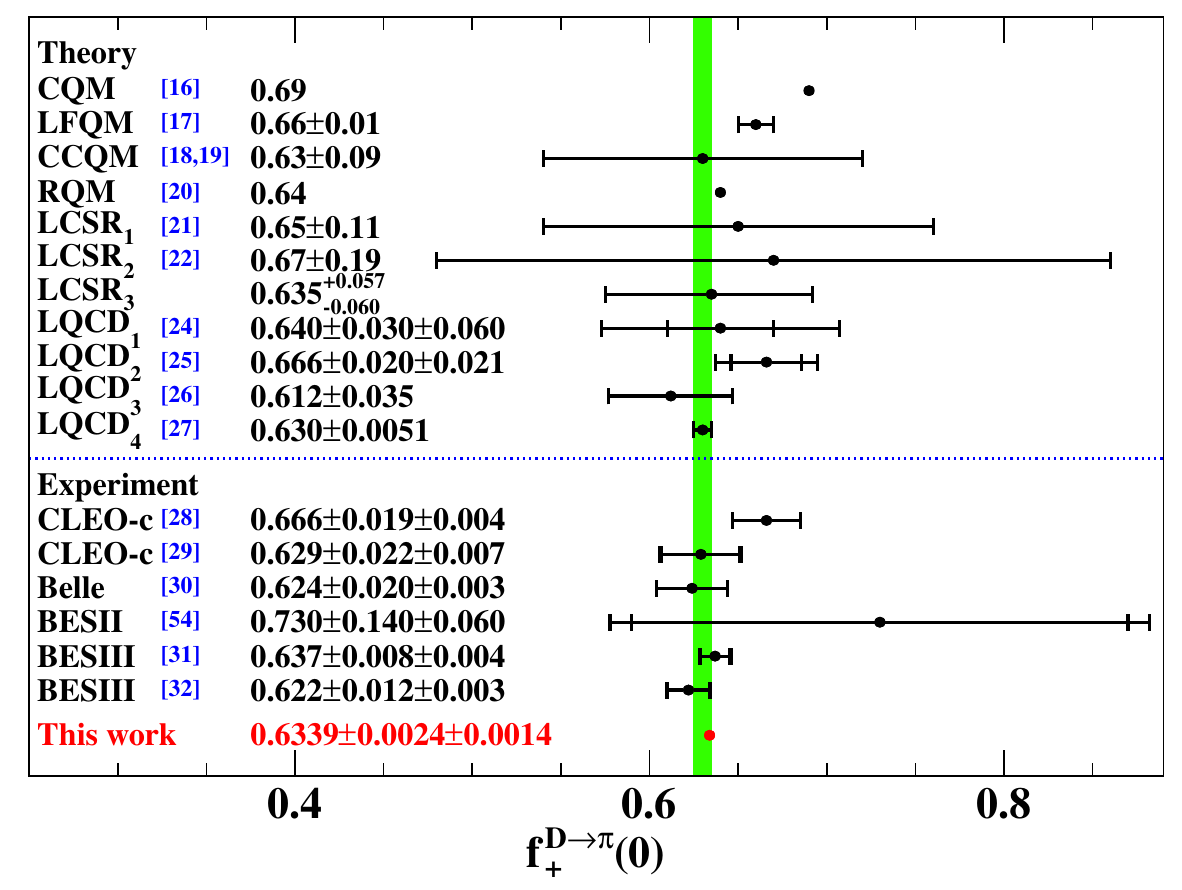}
	\caption{Comparison of the $f^{D\to\pi}_+(0)$ measured by this work with the theoretical calculations and other experiments.
		The green band corresponds to the $\pm1\sigma$ limit of the LQCD$_4$ calculation~\cite{FermilabLattice:2022gku}.
	}
	\label{compare_ff_pilnu}
\end{figure}

In summary, this Letter reports the precise BF measurements and decay dynamics analyses of $\pilnu$ decays. Based on the precisely measured full and PDRs, we examine $\mathcal{R}_{\rm LFU}^{0(+)}$ in full and separate $q^2$ intervals, and no evidence of LFU violation is found within the current data. By analyzing the $\pilnu$ decay dynamics without a SC, we obtain $\ffpi|V_{cd}|$.
Taking the $|V_{cd}|$ given by the PDG~\cite{pdg2024} as input, we obtain the hadronic FF $\ffpi = \fpi$. 
The comparison of the $\ffpi$ obtained in this work with the previous measurements and theoretical calculations is shown in Fig.~\ref{compare_ff_pilnu}.
Conversely, using the $\ffpi$ calculated in the LQCD~\cite{FermilabLattice:2022gku}, we obtain the CKM matrix element
$|V_{cd}| = \vcdpi$.
These results improve upon the previous best measurements~\cite{BESIII:2015tql,BESIII:2024kvt} by a factor of 2-3 in precision.
In addition, the $q^{2}$-binned angular observable forward-backward asymmetries $\mathcal{A}_{FB}$ and the
constraints on the complex Wilson coefficients (Re$(C_S^\mu)$ and Im$(C_S^\mu)$) in $D\to \pi\ell^+\nu_\ell$ are presented for the first time,
which benefit searches for new scalar bosons beyond the SM.

\textbf{Acknowledgement}

The BESIII Collaboration thanks the staff of BEPCII (https://cstr.cn/31109.02.BEPC) and the IHEP computing center for their strong support. This work is supported in part by National Key R\&D Program of China under Contracts Nos. 2023YFA1606000, 2023YFA1606704, 2025YFA1613900; National Natural Science Foundation of China (NSFC) under Contracts Nos. 11635010, 11935015, 11935016, 11935018, 12025502, 12035009, 12035013, 12061131003, 12192260, 12192261, 12192262, 12192263, 12192264, 12192265, 12221005, 12225509, 12235017, 12342502, 12361141819, 12535005; the Chinese Academy of Sciences (CAS) Large-Scale Scientific Facility Program; the Strategic Priority Research Program of Chinese Academy of Sciences under Contract No. XDA0480600; CAS under Contract No. YSBR-101; 100 Talents Program of CAS; The Institute of Nuclear and Particle Physics (INPAC) and Shanghai Key Laboratory for Particle Physics and Cosmology; Agencia Nacional de Investigación y Desarrollo de Chile (ANID), Chile under Contract No. ANID CCTVal CIA250027; ERC under Contract No. 758462; German Research Foundation DFG under Contract No. FOR5327; Istituto Nazionale di Fisica Nucleare, Italy; Knut and Alice Wallenberg Foundation under Contracts Nos. 2021.0174, 2021.0299, 2023.0315; Ministry of Development of Turkey under Contract No. DPT2006K-120470; National Research Foundation of Korea under Contract No. RS-2026-25486791; National Science and Technology fund of Mongolia; Polish National Science Centre under Contract No. 2024/53/B/ST2/00975; STFC (United Kingdom); Swedish Research Council under Contract No. 2019.04595; U. S. Department of Energy under Contract No. DE-FG02-05ER41374



\begin{thebibliography}{99}
\bibitem{supplement}
	See the joint submission at [URL will be inserted by publisher] for further details.
\end{thebibliography}
\end{document}